\documentclass[aps,twocolumn,showpacs,showkeys]{revtex4}
\usepackage{amsmath}
\usepackage{amssymb}
\usepackage[dvips]{graphicx}

\begin{document}

\title{Spinning Strings, Black Holes and Stable Closed Timelike Geodesics.}

\author{Val\'eria M. Rosa}
\email{e-mail: vmrosa@ufv.br}
\affiliation{
Departamento de Matem\'atica, Universidade Federal de Vi\c{c}osa, 
36570-000 Vi\c{c}osa, M.G., Brazil} 
\author{ Patricio S. Letelier}
\email{e-mail: letelier@ime.unicamp.br} 
 
\affiliation{
Departamento de Matem\'atica Aplicada-IMECC,
Universidade Estadual de Campinas,
13081-970 Campinas,  S.P., Brazil}

\begin{abstract}
  The existence and stability under linear perturbation of closed
  timelike curves in the spacetime associated to Schwarzschild black
  hole pierced by a spinning string are studied. Due to 
the superposition of the  black hole,
 we find that the spinning string spacetime is deformed in such a way
to allow the existence of  closed timelike geodesics. 
\end{abstract}

\keywords{
Closed Timelike Geodesics, Linear Stability,  Time Machines, Black Holes,
  Cosmic Strings,Torsion Lines}
\pacs{04.20.Gz, 04.20.Dg, 04.20.Jb}

\maketitle

The existance of  closed timelike curves (CTCs) in  the G\"odel universe and other 
apacetimes is a worrying fact since  these
curves show a clear  violation of causality. 
 In some cases these CTCs can be
 disregarded by energy considerations.
 Their existance requires an external force acting along
 the whole CTC, process that may consume a great amount of
 energy. The energy needed to travel along a  CTC in G\"odel's universe is
 computed in \cite{pfarr}.  When the external force is null 
 the energy needed to travel  is also null. Therefore, in principle, the
 existence of closed timelike geodesics (CTGs) 
presents a  bigger problem of breakdown of causality.

The classical problem of the existence of closed geodesics in
Riemannian geometry was solved by Hadamard~\cite{hada} in two
dimensions and by Cartan~\cite{cartan} in an arbitrary number of
dimensions.  As a topological problem, the existence of CTGs
 was proved by Tipler~\cite{tipler} in a
class of four-dimensional compact Lorentz manifolds with covering
space containing a compact Cauchy surface. In a compact pseudo-Riemaniann
manifold with Lorentzian signature (Lorentzian manifold)
Galloway~\cite{galloway} found sufficient conditions to have CTGs, see
also~\cite{guediri}.

To the best of our knowledge there are four solution to the Einstein
 equations generated by matter with positive mass density 
 that contain CTGs: a)
 Soares~\cite{soares} found a class of cosmological models, solutions
 of Einstein-Maxwell equations, with a subclass where the timelike
 paths of  matter are closed. For these models the existence of CTGs
 is demonstrated and explicit examples are given. 
These  CTGs are not linearly  stable \cite{other}. b)
 Steadman~\cite{steadman} described the behavior of CTGs in a vacuum
 exterior for the van Stockum solution that represents an infinite rotating dust
 cylinder. For this solution explicit examples of CTCs and CTGs are
 shown. There are stable CTGs in this spacetime \cite{other}. c) Bonnor and 
Steadman~\cite{bonnor} studied the existence of
 CTGs in a spacetime with two spinning particles each one with
 magnetic moment equal to angular moment and mass equal to charge
 (Perjeons), in particular, they present a explicit CTG. 
 This particular CTG is not stable, but there exist many other that
 are stable~\cite{rosalet1}. d)  There are   linearly 
stable CTGs \cite{other} in one of the
  G\"odel-type metrics with not flat background studied by G\"urses et al. 
 \cite{gurses}\cite{gleiser}.  For CTGs in a spacetime 
associated to a  cloud of strings with negative mass density see~\cite{gron}. These
 CTGs are not stable \cite{other}.

The existence of CTCs in a spacetime whose source is a spinning string
has been investigated by many authors (see for instance
\cite{deser}-\cite{soleng}). The interpretation of these strings as
torsion line defects can be found in \cite{letelier},  \cite{tld}, see also
\cite{sousa}\cite{petti}. These torsion line defects  appear when
one tries to stabilize two rotating black holes kept apart by spin
repulsion \cite{letol}.  Also, the black hole thermodynamics
associated to a static black hole pierced by a non rotating string was
studied some time ago by Aryal et al.~\cite{aryal}.

In the present work we study the existence and stability of CTCs under linear
perturbations in the spacetime associated to Schwarzschild
black hole (BH) pierced by a spinning string.  Even though this spacetime is 
more  a mathematical curiosity than an example of a real spacetime we believe 
that the study of stability of CTCs and CTGs can shed some light into 
the existence of this rather pathological curves.
In particular, we  study  sufficient conditions to have linearly stable CTGs. 
We find that these conditions are not very restrictive and can be easily 
satisfied.  Furthermore, we compared them with the
same conditions studied by Galloway \cite{galloway} for a compact
Lorentzian manifold.

Let us consider the spacetime with metric,
\begin{equation}
ds^2=(1-\frac{2m}{r})(dt -\alpha d\varphi)^2-\frac{dr^2} {1-
\frac{2m}{r}}-r^2(d\theta^2+\beta^2\sin^2\theta d\varphi^2),
\label{metric}
\end{equation}
where $\alpha=4S$ and $S$ is the string's spin  angular momentum per unit of length,
$\beta=1-4\lambda$ and $\lambda$ is the string's linear mass density  that is equal 
to its tension ($\lambda\leq 1/4$).

In the particular case, $\alpha=0$ and $\beta=1,$ the metric
 (\ref{metric}) reduces to the Schwarzschild solution. When $m=0,$
  ~Eq. (\ref{metric}) represents a spinning string, with the further
 specialization $\beta=1$ (not deficit angle) we have a pure massless torsion
 line defect \cite{letelier} \cite{tld}.  Therefore the metric
 (\ref{metric}) can be considered as representing the spacetime
 associated to a Schwarzschild black hole pierced by a spinning
 string.

Let us denote by $\gamma$ a closed curve given in its parametric form by, 
\begin{equation}
  t = t_0, \;\;
  r = r_0, \;\;
  \varphi \in [0,2\pi], \;\; \theta=\dfrac{\pi}{2}, 
\label{CTC}
\end{equation}
where $t_0$ and $r_0$ are constants. When $\gamma$ is parametrized  with 
an arbitrary parameter
 $\sigma$, we have a timelike curve when
 $\frac{dx^\mu}{d\sigma}\frac{dx_\mu}{d\sigma}>0$. This
 condition reduces to    $g_{\varphi\varphi} > 0$, i.e.,
\begin{equation}
(1-2m/r_0)\alpha^2-r_0^2\beta^2>0.
\label{tl_cond}
\end{equation}

A generic CTC $\gamma$ satisfies the system of equations given by 
\begin{equation}
\ddot{x}^\mu+\Gamma^{\mu}_{\alpha \beta}\dot{x}^\alpha\dot{x}^\beta  = F^\mu(x),
\label{CTCsystem}
\end{equation} where the overdot indicates
 derivation with respect to $s,\;$ $\Gamma^{\mu}_{\alpha \beta}$ are the 
Christoffel symbols and  $F^\mu$ is a 
specific external force  $(a^\mu=F^\mu).$ 
The nonzero component of the four-acceleration of $\gamma$ is 
\begin{equation}
  \label{acceleration}
a^r=\dfrac{1}{r_0^3}(r_0-2m)(\alpha^2m-r_0^3\beta^2)\dot{\varphi}^2.
\end{equation}

Our goal is to study the behavior of closed timelike geodesics. Therefore
taking $\alpha$ as one of the two solutions of
\begin{equation}
\alpha^2m-r_0^3\beta^2=0,
\label{alpha}
\end{equation}
 we have $a^r=0$. Under this
condition (\ref{tl_cond}) is satisfied when $r_0>3\,m$, that put the CTG
 outside the black hole.

Let $\tilde{\gamma}$ be the curve obtained from $\gamma$ after a small
perturbation ${\bf \xi}$,
i.e., $\tilde{x}^{\mu}=x^{\mu}+\xi^{\mu}$. 
 From equations (\ref{CTCsystem}) one finds that the system of differential
 equations satisfied by the perturbation ${\bf \xi}$ is~\cite{rosalet2},
\begin{equation}
\frac{d^2\xi^{\alpha}}{ds^2}+2\Gamma^{\alpha}_{\beta 
\mu}\frac{d\xi^{\beta}}{ds}u^{\mu}+\Gamma^{\alpha}_{\beta \mu,\lambda}\xi^{
\lambda}u^{\beta}u^{\mu}=F^{\alpha}_{,\lambda}\xi^{\lambda}.
\label{systemPerturbation}
\end{equation}

For the above mentioned closed timelike geodesic the
system~(\ref{systemPerturbation}) reduces to
\begin{eqnarray}
&&\ddot{\xi}^0+k_1\dot{\xi}^1=0,\\
\label{systemCTG_t}
&&\ddot{\xi}^1+k_2\dot{\xi}^0+k_3\xi^1=0,\\
\label{systemCTG_r}
&&\ddot{\xi}^2+k_4\dot{\xi}^1=0,\\ 
\label{systemCTG_phi}
&&\ddot{\xi}^3+k_5\xi^3=0,
\label{systemCTG_z}
\end{eqnarray}
where 
\begin{eqnarray}
&& k_1=2\Gamma^0_{21}\dot{\varphi},\, k_2=2\Gamma^1_{20}\dot{\varphi},\,
 k_3=\Gamma^1_{22,1}\dot{\varphi}^2, \nonumber \\
&& k_4=2\Gamma^2_{21}\dot{\varphi}, \, k_5=\Gamma^3_{22,3}\dot{\varphi}^2.
\label{genk}
\end{eqnarray}

 A curve  $\gamma$ parametrized by the proper time, $s$, is timelike
when  $\dot{x}^{\mu} \dot{x}_{\mu} = 1 $. For the curve  $\gamma(s)$ we have that
 this last condition gives us,
\begin{equation}
\dot{\varphi}^2=\frac{m}{\beta^2r_0^2(r_0-3m)}.
\end{equation}

The solution of (\ref{systemCTG_t})-(\ref{systemCTG_z})  is
\begin{equation}
\begin{array}{l}
\xi^0=-k_1(c_3\sin(\omega s+c_4)/\omega+\lambda s)+c_1\,s+c_5,\\ 
\xi^1=c_3\cos(\omega s+c_4)+\lambda, \\
\xi^2=-k_4(c_3\sin(\omega s+c_4)/\omega+\lambda s)+c_2\,s+c_6,\\ 
\xi^3=c_7\cos(\sqrt{k_5}s+c_8),
\end{array}\label{exactpert}
\end{equation}
where $c_i,\;i=1,\dots,8$ are integration constants,
\begin{eqnarray}
\omega&=&\sqrt{k_3-k_1k_2} \nonumber\\
&=&[\beta^2(r_0-6m)\dot{\varphi}^2/r_0]^{1/2}, 
\label{omega}
\end{eqnarray}
and $\lambda
= -k_2c_1/\omega^2$. Thus   when $r_0>6\,m$,  the constant
 $\omega$ is real and the solution  (\ref{exactpert}) shows  the 
typical behavior  for    stability, i.e.,  vibrational modes untangled 
with translational ones that can be 
eliminated by a suitable choice of the initial conditions.

When the black hole is removed,  we are left with the spacetime
 of the spinning string whose line
element is,
\begin{equation}
ds^2=(dt -\alpha d\varphi)^2-dr^2-r^2(d\theta^2+
\beta^2\sin^2\theta d\varphi^2).
\label{metric3}
\end{equation}
The closed curve, $\gamma$, is timelike when $\alpha^2-r_0^2\beta^2>0$. The
$a^r$-component of the four-acceleration is given by
$a^r=-\beta^2r_0\dot{\varphi}^2$. Thus for $r<|\alpha/\beta|$ we
 have closed timelike curves, which are not geodesics.

For the closed curve (\ref{CTC}) the
system~(\ref{systemPerturbation}) is written now as in
(\ref{systemCTG_t})-(\ref{systemCTG_z})  replacing equation
(\ref{systemCTG_r}) by
\begin{equation}
\ddot{\xi}^1+k_2\dot{\xi}^2+k_3\xi^1=
\partial_r(\Gamma^1_{22}\dot{\varphi}^2)\xi^1, \label{xi2}
\end{equation}
where now  $k_2=2\Gamma^1_{22}\dot{\varphi}$ and
$\dot{\varphi}^2=(\alpha^2-r_0^2\beta^2)^{-1}$.
In this particular case the solution of (\ref{systemPerturbation})
has the same form that (\ref{exactpert}) with
$\omega^2=2\beta^2\dot{\varphi}^2(2+\beta^2 r_0^2\dot{\varphi}^2)$. Therefore,
 the 
CTCs are stable.

In summary,  there exist linearly stable CTCs in the
spacetime related to a spinning string and these curves are restricted
to a small region of the spacetime. Closed timelike geodesics do not
exist in this spacetime.

 For the nonlinear superposition of a spinning string with a
 Schwarzschild black hole the new spacetime has linearly stable
 CTGs. The region of stability is the same of the usual
  circular geodesics in the  Schwarzschild black hole alone. 
The presence of the spinning string
 does not affect the stability of the orbits. It seems that torsion
 lines defects superposed to matter (not strings, $\beta=1$) is a
 main ingredient to have stable CTGs.  Loosely speaking, we have
 that a torsion line defect alone makes possible  the existence of  CTCs.  When the
 black hole is present the spinning string spacetime is deformed in such a way 
to  allow the   existence of a  CTG. 
This fact is also confirmed in
 the case of the two Perjeons solutions studied in~\cite{bonnor}
 wherein the torsion line defect is a main ingredient to have CTCs and
 CTGs.

It is instructive to look the previous results in a more direct and graphic way. 
The length of CTC in (\ref{CTC}) only depends on the value of $r=r_0$. We find,
\begin{eqnarray}
s(r_0)&=&2\pi\,\sqrt{g_{\varphi\varphi}(r_0)}, \nonumber \\
&=&2\pi [(1-2m/r_0)\alpha^2-r_{0}^2\beta^2]^{1/2}.
\label{s(r)}
\end{eqnarray}
 This function has a local maximum for 
\begin{equation}
r_m=(m\alpha^2/\beta^2)^{1/3}. \label{alpha2}
\end{equation}
Note that this equation is equivalent to (\ref{alpha}), the
condition to have a geodesic.

The role of the black hole mass,  in the appearance of  CTGs, is 
to produce a local maximum
 in the length function, $s(r_0)$. This maximum gives us the position  of  the CTG
that in our case is located  outside of the source of the spacetime,  beyond  
the black hole horizon.

\begin{figure}
\includegraphics[scale=.5]{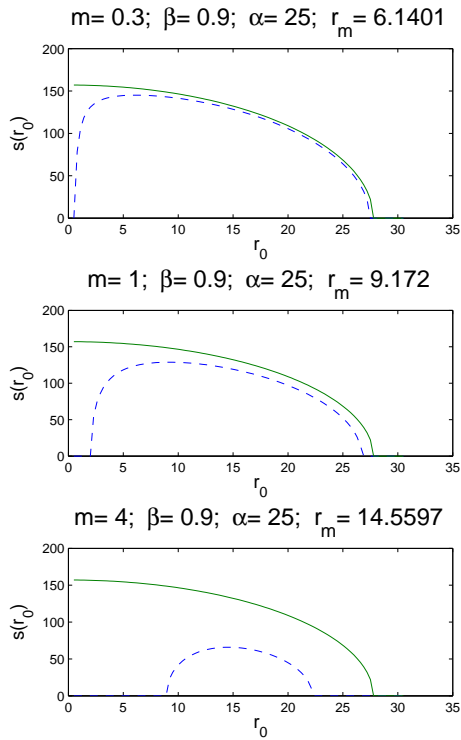}
\caption{The function $s(r_0)$ for a spinning string (solid line) 
and for a black hole pierced by the string (dashed line). We see 
how the presence of the mass shift the maximum of $s(r_0)$ for the 
string that is located at $r_0=0$ to a position outside the black
 hole horizon. The maximum,
$r_m$ represent the radius of the CTGs, the first two are stable
 and the second is  not.}
\label{fig1}
\end{figure}

In Fig. 1 we present, as a solid line the function $s(r_0)$ for a spinning string,
and as  a dashed line the same function for the superposition of the black hole with
the previously mentioned string for the same values of the  parameters 
$\alpha=25$ (spin parameter) and deficit angle parameter $\beta=0.9$ and  
different values of the black hole mass ($m=0.3,1,4$).
We see how the presence 
of the mass shift the maximum for the string located at $r_0=0$ to a position 
$r_0>3m$. Also the points under the curves represent the pairs
$[r_0,s(r_0)]$ for CTCs in each case. We note that the region for CTCs  for 
the black hole pierced by the string diminishes when the mass 
increases. The maximum of the dashed line
represents the CTG. We see, that in the first two cases the CTGs are 
stable ($r_m>6m$) and in the last case the CTG is not stable ($r_m<6m$).
\begin{figure}
\includegraphics[scale=.5]{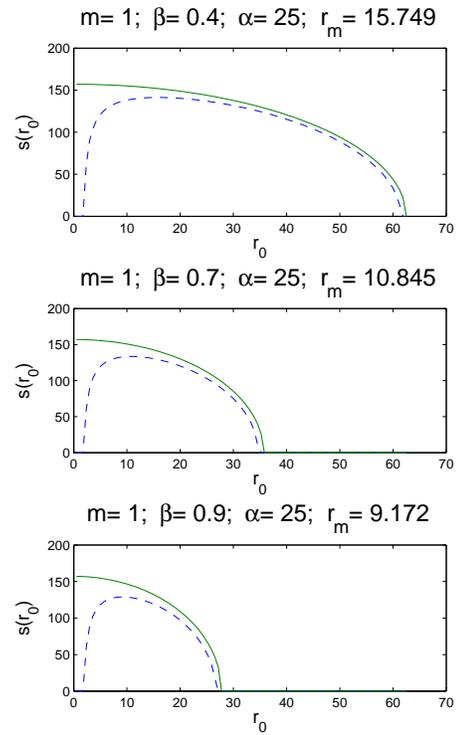}
\caption{The function $s(r_0)$ for a spinning string (solid line) and for 
a black hole pierced by the string (dashed line). We see how the size of
 the deficit angle parameter 
$\beta$ changes the region for CTCs and the value of $r_m.$ }
\label{fig2}
\end{figure}  

In  Fig. 2 we keep the value of the black hole mass constant, $m=1$, as
 well as, the spin parameter, $\alpha=25,$ and change the deficit angle 
parameter $\beta =0.4,0.7,0.9.$ We see that
the larger the string density, $\lambda=(1-\beta)/4,$ the larger the region for CTCs.

\begin{figure}
\includegraphics[scale=.5]{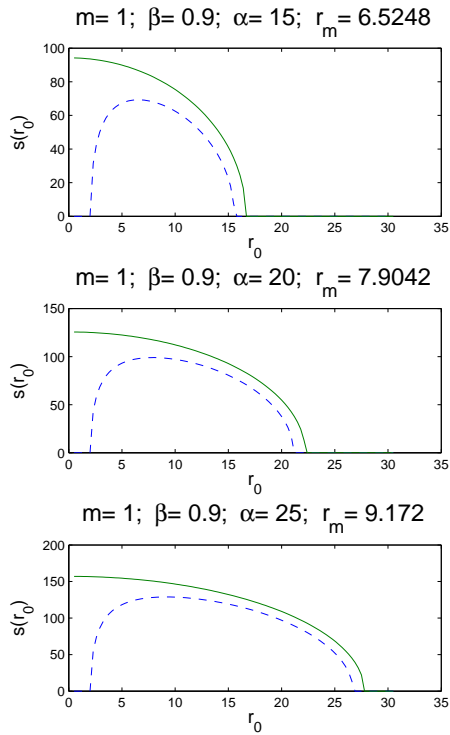}
\caption{The function $s(r_0)$ for a spinning string (solid line) 
and for a black hole pierced by the string (dashed line). We see how
 the size of the spin parameter 
$\alpha$ changes the region for CTCs and the value of $r_m$. The spin
 parameter, in this case,  is the essential ingredient to have CTCs and CTGs.}
\label{fig3}
\end{figure}  

In  Fig. 3 we keep the value of the black hole mass constant, $m=1$, as well as, 
the  deficit angle parameter $\beta = 0.9$  and change the spin 
parameter $\alpha=15, 20, 25$.
We see that  the regions where the CTCs appear are larger for bigger spin 
parameter. This parameter is essential to have CTCs and CTGs in this case.

As we said before the existence of a CTGs does not put restrictions on the
energy to travel along this curve. Furthermore,   the force needed to move 
near a stable geodesic 
is  small.  Therefore,  the energy required  will be also small. In principle
this small force can be provided by and engine, say a rocket. Hence  there will be not a 
severe  energy restriction  to travel near to a geodesic. Furthermore, when moving along a 
 stable  CTG   the control problem is a trivial one. Small trajectory corrections require small
energy, also we do not have the danger to enter into a   run away situation.

A result from Galloway~\cite{galloway} states that in a compact
Lorentzian manifold, each stable free t-homotopy class contains a
longest closed timelike curve, and this curve is necessarily a closed
timelike geodesic. The assumption that $M$ be compact can be
weakened, it is sufficient to assume that there exists an open set $U$
in $M$ with compact closure such that each curve $\gamma \in {\mathcal C}$ 
(the free t-homotopy class) is contained in $U$. In our case th G\"odel universe and other 
apacetimes e region 
containing the CTCs in $\mathcal C$ is not compact. Therefore Galloway's
conditions do not apply in this case, they too strong.

We want to point out that the
stability of the circular orbits does not depend on the fact of the
orbit be a  CTG. We found the same region of stability of the
usual circular geodesics.  This result is not surprising since
 our pierced black hole is locally identical to a usual black
 hole. Moreover  one can consider
black holes surrounded by  different axially symmetric distributions of
 matter \cite{orbits} pierced by a spinning string. 
In  this case,
 depending on the different parameters of the solution,  we can 
also have CTGs and their stability will be the same as the usual 
 circular orbits considered in \cite{orbits}.

Furthermore, we analyze if the CTGs studied in the present work
satisfy the sufficient conditions of Galloway's theorem for the
existence of CTGs. We found that ours CTGs do not satisfy these
conditions. The possibility of an example that satisfy exactly the
conditions of this theorem is under study.  We want to mention that
the solution of Einstein equations considered in this work is much
simpler that the  ones listed in the introduction.

Finally, we notice  that the  spacetime associated to the black 
hole pierced by a spinning string  is not a counter 
example to the Chronology Protection Conjecture~\cite{cpc} that
essentially says that the laws of the physics do not allow the
appearance of closed timelike curves.  A valid dynamic to built
this spacetime  is not known.

\vspace*{0.3cm}

V.M.R.  thanks Departamento de Matem\'atica-UFV for giving the
conditions to finish this work which was partially supported  by
PICDT-UFV/CAPES. P.S.L. thanks the partial financial
support of FAPESP and CNPq.


\end{document}